\documentclass[aps,preprint,showpacs,preprintnumbers,amsmath,amssymb]{revtex4}
\usepackage{amsmath,mathrsfs,amsbsy,color,graphicx,bm,amsthm,amsfonts}
\usepackage{units}
\usepackage{bbm}
\usepackage{times}
\usepackage{dcolumn}
\usepackage{mathrsfs}
\usepackage{amsmath,amssymb,epsfig}
\usepackage{amsmath}
\usepackage{float}
\newcommand{\udots}{\mathinner{\mskip1mu\raise1pt\vbox{\kern7pt\hbox{.}}
\mskip2mu\raise4pt\hbox{.}\mskip2mu\raise7pt\hbox{.}\mskip1mu}}
\begin{document}
\title{Genuinely accessible and inaccessible entanglement  in Schwarzschild black hole }
\author{Shu-Min Wu$^1$\footnote{Email: smwu@lnnu.edu.cn}, Xiao-Wei Teng$^1$, Jin-Xuan Li$^1$, Si-Han Li$^1$, Tong-Hua Liu$^2$\footnote{Corresponding author: liutongh@yangtzeu.edu.cn}, , Jie-Ci Wang$^3$\footnote{Corresponding author: jcwang@hunnu.edu.cn}}
\affiliation{$^1$ Department of Physics, Liaoning Normal University, Dalian 116029, China\\
$^2$ Department of Physics, Yangtze University, Jingzhou 434023, China\\
$^3$ Department of Physics, Hunan Normal University, Changsha 410081, China}


\begin{abstract}
The genuine entanglement of Dirac fields for an N-partite system is investigated in Schwarzschild spacetime and the analysis is carried out using the single-mode approximation. Due to the Hawking effect, quantum entanglement is divided into two parts physically accessible and inaccessible entanglement. We obtain a general analytic expression of genuine N-partite entanglement that includes all accessible and inaccessible entanglement in a Schwarzschild black hole. Unlike bosonic entanglement, the accessible N-partite entanglement of Dirac fields monotonically decreases to a nonzero value with the Hawking temperature.
Interestingly, the inaccessible N-partite entanglement is a monotonic or non-monotonic function of the Hawking temperature, depending on the ratio between accessible and inaccessible  modes, in contrast to bipartite or tripartite entanglement that is only a monotonic function of the Hawking temperature. Finally, we obtain two restrictive relationships for the quantum information of the black hole. This conclusion provides a new understanding of Hawking effect of the black hole.
\end{abstract}

\vspace*{0.5cm}
 \pacs{04.70.Dy, 03.65.Ud,04.62.+v }
\maketitle
\section{Introduction}
General relativity first predicted black holes, and then astronomy indirectly proved the existence of black holes \cite{L24,L25,L26,L27,L28,L29,L30,L31}. Although the study of black holes has made great progress, due to the special properties of black holes and their  distant distance from us, the mystery of black holes has not yet been solved. Therefore, so far, the research on black holes has mostly remained theoretical. From the classical point of view, a particle falls into the event horizon of a black hole, the event horizon prevents the particle from returning, and escape becomes impossible. However, given quantum effects, the particles inside the black hole are destined to gradually escape to the outside, resulting in  Hawking radiation \cite{L32}.
This phenomenon is an intermediate bridge between quantum mechanics and gravity, and is at the heart of the information paradox of black hole \cite{L33,L34,L35}.
In particular, quantum entanglement influenced by Hawking radiation is a possible important way to solve the information paradox of the black hole.

Quantum information in the gravitational background, which is a combination of quantum information, quantum field theory, and gravity,  mainly involves two aspects of research: (i) it can use quantum technology to probe the structure of spacetime; (ii) it can explore how gravitational effects affect quantum resources. Quantum information in the gravitational background has received extensive attention in theory, simulation, and experiment.
Theoretically, the influence of the Hawking effect of the black hole on quantum steering, entanglement, and discord has been studied extensively \cite{L38,L39,L40,L41,L42,L43,L44,L45,L46,L47,L48,L49,L50,L51,L52,L53,L54,L55,L56,L57,QLQ57,L80}.
Simulationally, some schemes  have attempted to simulate Hawking radiation in a dc-SQUID array transmission line, a Bose-Einstein condensate,  an optical analogue, and a superconducting circuit \cite{A1,A2,A3,A4,A5}. Experimentally,  Pan and his team have shown quantum decoherence effect caused by the gravitational effect of the Earth through the "Micius", marking that the research on quantum information in the gravitational background has entered the stage of precision experimental verification \cite{A6}. With the increasing complexity of quantum information tasks, we need to use multipartite entanglement to deal with relativistic quantum information tasks. Therefore, we need to study the behavior of N-partite entanglement in the gravitational background.
In addition, we try to explore quantum information of the black hole with genuine entanglement in multipartite systems.

In this paper, we will investigate the properties of genuine N-partite entanglement of Dirac fields and its restrictive relationships in the background of the Schwarzschild black hole.
Assume that $N$ Kruskal observers initially share a
Greenberger-Horne-Zeilinger-like state at the asymptotically flat region. Next, we let $n$ ($n<N$) observers hover near the event horizon of the black hole, while the rest $N-n$  Kruskal observers remain at the asymptotically flat region. Due to the Hawking effect of the black hole, $n$ imagined observers are segregated by the event horizon. We attempt to obtain a concise analytical expression including all physically accessible and inaccessible genuine N-partite entanglement in Schwarzschild spacetime. We will compare genuine N-partite entanglement with bipartite entanglement (or tripartite entanglement) and get completely different behaviors between them in curved spacetime. Finally, we will obtain two restrictive relationships between accessible and inaccessible N-partite entanglement, showing the flow  of multipartite quantum information of the black hole.

The paper is organized as follows. In Sec. II, we introduce the quantization of Dirac field in the background of Schwarzschild black hole.  Sec. III, we study the Hawking effect of the black hole on  physically accessible and inaccessible genuine N-partite entanglement.
Finally, the last section is devoted to the conclusion.

\section{Quantization of Dirac field in  Schwarzschild black hole  \label{GSCDGE}}
The metric of the Schwarzschild black hole becomes
\begin{eqnarray}\label{S7}
ds^2&=&-(1-\frac{2M}{r}) dt^2+(1-\frac{2M}{r})^{-1} dr^2\nonumber\\&&+r^2(d\theta^2
+\sin^2\theta d\varphi^2),
\end{eqnarray}
where $M$ is the mass of the Schwarzschild black hole \cite{L42}. We set $G$, $c$, $\hbar$ and $k_B$ as unity in this paper.  The massless Dirac equation $[\gamma^a e_a{}^\mu(\partial_\mu+\Gamma_\mu)]\Phi=0$  in Schwarzschild spacetime
can be specifically expressed in the following form
\begin{eqnarray}\label{S9}
&-&\frac{\gamma_{0}}{\sqrt{1-\frac{2M}{r}}}\frac{\partial\Phi}{\partial t}+\gamma_{1}\sqrt{1-\frac{2M}{r}}[\frac{\partial}{\partial r}+\frac{1}{r}\\ \nonumber&+&\frac{M}{2r(r-2M)}]\Phi
+\frac{\gamma_{2}}{r}(\frac{\partial}{\partial\theta}+\frac{\cot\theta}{2})\Phi
+\frac{\gamma_{3}}{r\sin\theta}\frac{\partial\Phi}{\partial\varphi}=0,
\end{eqnarray}
where $\gamma_{i}$ (i = 0, 1, 2, 3) are the Dirac matrices \cite{Q56,Q57}. After solving Eq.(\ref{S9}), we get the positive fermionic frequency outgoing solutions
\begin{eqnarray}\label{S10}
\Phi^+_{{k},{\rm out}}\sim \phi(r) e^{-i\omega u},
\end{eqnarray}
\begin{eqnarray}\label{S11}
\Phi^+_{{k},{\rm in}}\sim \phi(r) e^{i\omega u},
\end{eqnarray}
where $\phi(r)$ is the four-component Dirac spinor and $u=t-r_{*}$ with the tortoise coordinate $r_{*}=r+2M\ln\frac{r-2M}{2M}$.
Therefore, we can expand Dirac field $\Phi$ through Eqs.(\ref{S10}) and (\ref{S11}) as
\begin{eqnarray}\label{S12}
\Phi&=&\int dk[\hat{a}^{in}_{k}\Phi^+_{{k},{in}}+\hat{b}^{in\dag}_{-k}\Phi^-_{{-k},{ in}}\\ \nonumber&+&\hat{a}^{out}_{k}\Phi^+_{{k},{ out}}+\hat{b}^{out\dag}_{-k}\Phi^-_{{-k},{ out}}],
\end{eqnarray}
where $\hat{a}^{in}_{k}$ is the fermionic annihilation operator inside the event horizon, and $\hat{b}^{out\dag}_{-k}$ is the antifermionic creation operator outside the event horizon in the Schwarzschild black hole. Note that the modes $\Phi^\pm_{{k},{in}}$ and $\Phi^\pm_{{k},{out}}$ are usually called Schwarzschild modes.

Using Domour and Ruffini's suggestions \cite{L32}, one can introduce a complete basis for the positive energy mode (Kruskal mode)  to connect Eqs.(\ref{S10}) and (\ref{S11})
\begin{eqnarray}\label{S13}
\Psi^+_{{k},{out}}=e^{-2\pi M\omega}\Phi^-_{{-k},{in}}+e^{2\pi M\omega}\Phi^+_{{k},{out}},
\end{eqnarray}
\begin{eqnarray}\label{S14}
\Psi^+_{{k},{in}}=e^{-2\pi M\omega}\Phi^-_{{-k},{out}}+e^{2\pi M\omega}\Phi^+_{{k},{in}}.
\end{eqnarray}
Therefore, the Kruskal modes can also be used to expand the Dirac field
\begin{eqnarray}\label{S15}
\Phi&=&\int dk[2\cosh(4\pi M\omega)]^{-\frac{1}{2}}[\hat{c}^{in}_{k}\Psi^+_{k,in}\\ \nonumber&+&\hat{d}^{in\dagger}_{-k}\Psi^-_{-k,in}+\hat{c}^{out}_{k}\Psi^+_{k,out}+\hat{d}^{out\dagger}_{-k}\Psi^-_{-k,out}],
\end{eqnarray}
where $\hat{c}^{\sigma}_{k}$ and $\hat{d}^{\sigma\dagger}_{k}$ with $\sigma=(in,out)$ are the fermionic annihilation operators and the antifermionic creation operators acting on the Kruskal vacuum.

Eqs.(\ref{S12}) and (\ref{S15}) indicate that the
Dirac field can be quantized by Schwarzschild and Kruskal modes, respectively, resulting in the Bogoliubov transformations between Schwarzschild and Kruskal operators, which  take the forms
\begin{eqnarray}\label{S16}
\hat{c}^{out}_{k}=\frac{1}{\sqrt{e^{-8\pi M\omega}+1}}\hat{a}^{out}_{k}-\frac{1}{\sqrt{e^{8\pi M\omega}+1}}\hat{b}^{in\dagger}_{-k},
\end{eqnarray}
\begin{eqnarray}\label{S17}
\hat{c}^{out\dagger}_{k}=\frac{1}{\sqrt{e^{-8\pi M\omega}+1}}\hat{a}^{out\dagger}_{k}-\frac{1}{\sqrt{e^{8\pi M\omega}+1}}\hat{b}^{in}_{-k}.
\end{eqnarray}
Employing Bogoliubov transformations, the expressions of the Kruskal vacuum and excited states in Schwarzschild spacetime read \cite{L46,Q57}
\begin{eqnarray}\label{ZS18}
\nonumber |0\rangle_K&=&\frac{1}{\sqrt{e^{-\frac{\omega}{T}}+1}}|0\rangle_{out} |0\rangle_{in}+\frac{1}{\sqrt{e^{\frac{\omega}{T}}+1}}|1\rangle_{out} |1\rangle_{in},\\
|1\rangle_K&=&|1\rangle_{out} |0\rangle_{in},
\end{eqnarray}
where $T=\frac{1}{8\pi M}$ is the Hawking temperature, ${|n\rangle_{out}}$  and ${|n\rangle_{in}}$ indicate the fermionic modes  outside the event horizon and the antifermionic
modes inside the event horizon, respectively.
\section{Genuinely accessible and inaccessible N-partite entanglement in Schwarzschild black hole  \label{GSCDGE}}
Genuine N-partite entanglement is an important resource with obvious advantages in quantum information tasks, such as multipartite quantum networks, quantum computing by cluster states, measurement-based quantum computation, and high sensitivity in metrology tasks \cite{L14,L16,L17,L18,L19,L20,L21,L22,L23}. We briefly introduce concurrence to measure genuine N-partite entanglement.
For the N-partite systems, the Hilbert-space orthonormal bases can be defined as $\{|0,0,...,0\rangle,|0,0,...,1\rangle,...,|1,1,...,1\rangle\}$. The X-state of the N-partite system is expressed in terms of density matrix as
\begin{eqnarray}\label{Q12}
 \rho_X= \left(\!\!\begin{array}{cccccccc}
A_1 &  &  &  &  &  &  & C_1\\
 & A_2 &  &  &  &  & C_2 & \\
 &  & \ddots &  &  &  \udots &  & \\
 &  &  & A_n & C_n &  &  & \\
 &  &  & C_n^* &  B_n & &  & \\
 &  & \udots &  &  & \ddots &  & \\
 & C_2^* &  &  &  &  & B_2 & \\
C_1^* &  &  &  &  &  &  & B_1
\end{array}\!\!\right),
\end{eqnarray}
where $n=2^{N-1}$. This conditions $\sum_i(A_i+B_i)=1$ and $|C_i|\leq\sqrt{A_iB_i}$ are  to ensure that $\rho_X$ is normalized and positive.
The genuine concurrence for the N-partite X-state is found to be
\begin{eqnarray}\label{Q13}
C(\rho_X)=2\max \{0,|C_i|-\mu_i \},   i=1,\ldots,n,
\end{eqnarray}
where $\mu_i=\sum_{j\neq i}^n\sqrt{A_jB_j}$ \cite{Q58}.

We  initially share an N-partite Greenberger-Horne-Zeilinger (GHZ) entangled state at the asymptotically flat region of the Schwarzschild black hole
\begin{eqnarray}\label{Q14}
|\psi\rangle_{1,\ldots,N}=\alpha|0\rangle^{\otimes N}+\sqrt{1-\alpha^2}|1\rangle^{\otimes N},
\end{eqnarray}
where $\alpha$ is a state parameter that runs from $0$ to $1$, and  the mode $i$ ($i = 1, 2, \ldots, N$) is observed by the observer $O_i$.
 After that,  we assume that $N-n$ observers remain at the asymptotically flat region, while $n$ ($n<N$) observers hover near the event horizon of the black hole.
Due to the Hawking effect, $n$ imagined observers are inside the event horizon of the black hole. From Eq.(\ref{ZS18}), we can see that the change from Kruskal modes to Schwarzschild
modes corresponds to the two-mode squeezing transformations.
Under such transformations, $N$ Kruskal modes are mapped into $N-n$ Kruskal modes, $n$ modes for the exterior region, and $n$ modes for the interior region of the black hole.  Therefore,  Eq.(\ref{Q14}) in terms of Schwarzschild modes can be rewritten as
\begin{eqnarray}\label{Q15}
|\psi\rangle_{1,\ldots,N+n}&=&\alpha\bigg[(\overbrace{|0\rangle_{n+1}|0\rangle_{n+2}...|0\rangle_{N}}^{|\bar 0\rangle})\bigotimes_{i=1}^{n}(\frac{1}{\sqrt{e^{-\frac{\omega}{T}}+1}}|0\rangle_{{\rm out,}i} |0\rangle_{{\rm in,}i}
+\frac{1}{\sqrt{e^{\frac{\omega}{T}}+1}} \\
\nonumber &\times&|1\rangle_{{\rm out,}i}|1\rangle_{{\rm in,}i})\bigg]+\sqrt{1-\alpha^2}\bigg[(\overbrace{|1\rangle_{n+1}|1\rangle_{n+2}...|1\rangle_{N}}^{|\bar 1\rangle})\bigotimes_{i=1}^{n}
(|1\rangle_{{\rm out,}i} |0\rangle_{{\rm in,}i})\bigg],
\end{eqnarray}
where $|\bar 0\rangle=|0\rangle_{n+1}|0\rangle_{n+2}...|0\rangle_{N}$
and $|\bar 1\rangle=|1\rangle_{n+1}|1\rangle_{n+2}...|1\rangle_{N}$.

If we take the trace over physically accessible $n-p$ modes and inaccessible $n-q$ modes of $|\psi\rangle_{1,\ldots,N+n}$, we obtain a general system $\rho_{N-n,p,q}$ ($p+q=n$), which
is composed of $N-n$ Kruskal modes, accessible $p$ modes for the exterior region, and  inaccessible $q$ modes for the interior region of the black hole.
In addition, we analyze the general density operator $\rho_{N-n,p,q}$ from the perspective of basis transformation. Eq.(\ref{ZS18}) leads to the evolution of the density operator from the Kruskal
basis to the Schwarzschild basis outside the event horizon as
\begin{eqnarray}
\nonumber Tr_{in}|0\rangle_K\langle0|&=&\frac{1}{{e^{-\frac{\omega}{T}}+1}}|0\rangle_{out} \langle0|+\frac{1}{{e^{\frac{\omega}{T}}+1}}|1\rangle_{out} \langle1|, \\Tr_{in}|0\rangle_K\langle1|&=&\frac{1}{\sqrt{e^{-\frac{\omega}{T}}+1}}|0\rangle_{out}\langle1|,\quad Tr_{in}|1\rangle_K\langle1|=|1\rangle_{out}\langle1|. \nonumber
 \end{eqnarray}
We trace over the modes supported in the interior region of the black hole as
\begin{eqnarray}
\nonumber Tr_{out}|0\rangle_K\langle0|&=&\frac{1}{{e^{-\frac{\omega}{T}}+1}}|0\rangle_{in} \langle0|+\frac{1}{{e^{\frac{\omega}{T}}+1}}|1\rangle_{in} \langle1|, \\Tr_{out}|0\rangle_K\langle1|&=&\frac{1}{\sqrt{e^{\frac{\omega}{T}}+1}}|1\rangle_{in}\langle0|,\quad Tr_{out}|1\rangle_K\langle1|=|0\rangle_{in}\langle0|. \nonumber
 \end{eqnarray}
Based on the above equations, we can easily find that the off-diagonal (diagonal) elements of the initial state given by Eq.(\ref{Q14}) are always on the off-diagonal (diagonal) elements of the X-state from the Kruskal basis to the Schwarzschild basis. Therefore,
the general density operator $\rho_{N-n,p,q}$ is still the X-state that is found to be
\begin{eqnarray}\label{Q22}
\rho_{N-n,p,q}=\boldsymbol{\rho_M}+\boldsymbol{\rho_V} +\boldsymbol{\rho_V}^\dagger +\boldsymbol{\rho_N},
\end{eqnarray}
with
\begin{eqnarray}
\nonumber\boldsymbol{\rho_M}&=&\frac{\alpha^2}{({e^{-\frac{\omega}{T}}+1})^n}|\bar0 \rangle\langle\bar0|\left[\bigotimes_{i=1}^{p}(|0\rangle_{{\rm out,}i}\langle0|)\right]\left[\bigotimes_{j=1}^{q}(|0\rangle_{{\rm in,}j}\langle0|)\right]+\frac{\alpha^2}{({e^{-\frac{\omega}{T}}+1})^{n-1}}\\ \nonumber
&\times&\frac{1}{{e^{\frac{\omega}{T}}+1}}|\bar0 \rangle\langle\bar0|\bigg\{\sum_{m=1}^p [(|1\rangle_{{\rm out,}m}\langle1|)\bigotimes_{i=1,i\neq m}^{p}(|0\rangle_{{\rm out,}i}\langle0|)\bigotimes_{j=1}^{q}(|0\rangle_{{\rm in,}j}\langle0|)]\\ \nonumber
&&\hspace{0.3cm} +\sum_{m=1}^q [\bigotimes_{i=1}^{p}(|0\rangle_{{\rm out,}i}\langle0|)(|1\rangle_{{\rm in,}m}\langle1|)\bigotimes_{j=1,j\neq m}^{q}(|0\rangle_{{\rm in,}j}\langle0|)]\bigg\}\\ \nonumber
&+&\frac{\alpha^2}{({e^{-\frac{\omega}{T}}+1})^{n-2}({e^{\frac{\omega}{T}}+1})^2}|\bar0 \rangle\langle\bar0|
\\ \nonumber && \hspace{0.3cm}\times\bigg\{\sum_{m=2}^p\sum_{z=1}^{m-1} [(|1\rangle_{{\rm out,}z}\langle 1|)(|1\rangle_{{\rm out,}m}\langle 1|)\bigotimes_{i=1,i\neq z,m}^{p}(|0\rangle_{{\rm out,}i}\langle0|)\bigotimes_{j=1}^{q}(|0\rangle_{{\rm in,}j}\langle0|)]
\\ \nonumber && \hspace{0.3cm}+\sum_{m=2}^q \sum_{z=1}^{m-1}[\bigotimes_{i=1}^{p}(|0\rangle_{{\rm out,}i}\langle0|)(|1\rangle_{{\rm in,}z}\langle1|)(|1\rangle_{{\rm in,}m}\langle1|)\bigotimes_{j=1,j\neq z,m}^{q}(|0\rangle_{{\rm in,}j}\langle0|)]
\\ \nonumber && \hspace{0.3cm}+\sum_{z=1}^p\sum_{m=1}^q [(|1\rangle_{{\rm out,}z}\langle1|)\bigotimes_{i=1,i\neq z}^{p}(|0\rangle_{{\rm out,}i}\langle0|)(|1\rangle_{{\rm in,}m}\langle1|)\bigotimes_{j=1,j\neq m}^{q}(|0\rangle_{{\rm in,}j}\langle0|)]\bigg\}\\ \nonumber
&+&...\\ \nonumber
&+&\frac{\alpha^2}{({e^{\frac{\omega}{T}}+1})^{n}}|\bar0 \rangle\langle\bar0|\bigotimes_{i=1}^{p}(|1\rangle_{{\rm out,}i}\langle1|)\bigotimes_{j=1}^{q}(|1\rangle_{{\rm in,}j}\langle1|),
\end{eqnarray}
\begin{eqnarray}
\nonumber\boldsymbol{\rho_V}=\frac{\alpha\sqrt{1-\alpha^2}}{\sqrt{({e^{-\frac{\omega}{T}}+1})^{p}({e^{\frac{\omega}{T}}+1})^q}}\bigg\{|\bar0 \rangle\langle\bar1|\bigotimes_{i=1}^{p}(|0\rangle_{{\rm out,}i}\langle1|)\bigotimes_{j=1}^{q}(|1\rangle_{{\rm in,}j}\langle0|)\bigg\},
\end{eqnarray}
and
\begin{eqnarray}
\nonumber\boldsymbol{\rho_N}=(1-\alpha^2)|\bar1 \rangle\langle\bar1|\bigotimes_{i=1}^{p}(|1\rangle_{{\rm out,}i}\langle1|)\bigotimes_{j=1}^{q}(|0\rangle_{{\rm in,}j}\langle0|).
\end{eqnarray}
From Eq.(\ref{Q22}), we can see that the  elements of the sub-density operators $\boldsymbol{\rho_M}$  and $\boldsymbol{\rho_N}$  are the diagonal elements, and  the  elements of the sub-density operator $\boldsymbol{\rho_V}$ are the off-diagonal elements. We can write the density operator $\rho_{N-n,p,q}$ as a simple density matrix that takes the matrix form
\begin{eqnarray}\label{Q23}
 \rho_{N-n,p,q}= \left(\!\!\begin{array}{cc}
 \mathcal{M} & \mathcal{V} \\
 \mathcal{V}^T & \mathcal{N} \\
 \end{array}\!\!\right),
\end{eqnarray}
in the $2^{n+1}$ basis $\{|\bar00...00\rangle,|\bar00...01\rangle,...,|\bar11...10\rangle,|\bar11...11\rangle \}.$
The sub-matrixes $\mathcal{M}$, $\mathcal{N}$, and $\mathcal{V}$ are detailed in Appendix A.
Substitution of the elements of the matrix $\rho_{N-n,p,q}$ into Eq.(\ref{Q12}), we can obtain the general genuine N-partite entanglement measured by the concurrence between $N-n$ Kruskal modes, accessible $p$ modes, and inaccessible $q$ modes as
\begin{eqnarray}\label{Q27}
C(\rho_{N-n,p,q})=\frac{2\alpha\sqrt{1-\alpha^2}}{\sqrt{({e^{-\frac{\omega}{T}}+1})^{p}({e^{\frac{\omega}{T}}+1})^q}} .
\end{eqnarray}
From Eq.(\ref{Q27}), we can find that the general genuine N-partite entanglement not only depends on the initial state parameters $\alpha$ but also on the $T$, $p$, and $q$. We note that the general entanglement  $C(\rho_{N-n,p,q})$ includes all accessible and inaccessible entanglement: (i) for $p=n$, the general genuine N-partite entanglement $C(\rho_{N-n,p,q})$
is an accessible  entanglement; (ii) for $p\neq n$, the general genuine N-partite entanglement $C(\rho_{N-n,p,q})$ is an inaccessible  entanglement.

\begin{figure}[H]
\centering
\includegraphics[height=1.8in,width=2.0in]{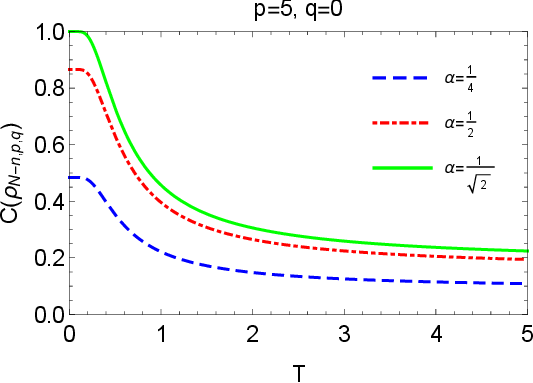}
\includegraphics[height=1.8in,width=2.0in]{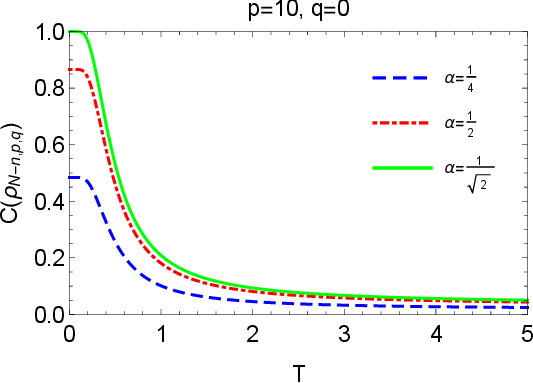}
\includegraphics[height=1.8in,width=2.0in]{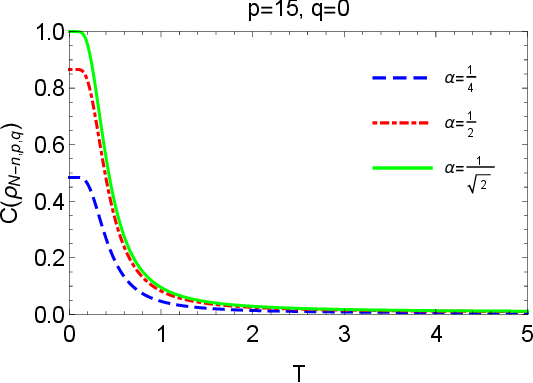}

\includegraphics[height=1.8in,width=2.0in]{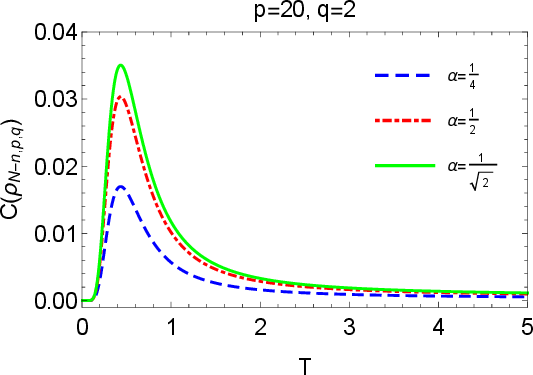}
\includegraphics[height=1.8in,width=2.0in]{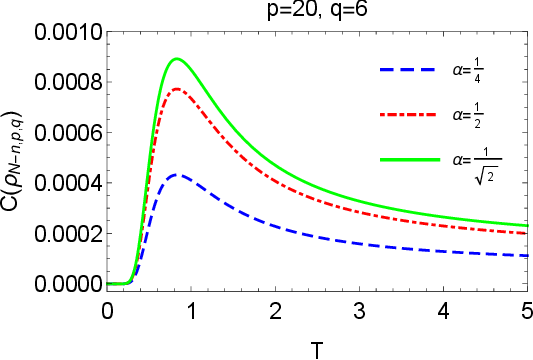}
\includegraphics[height=1.8in,width=2.0in]{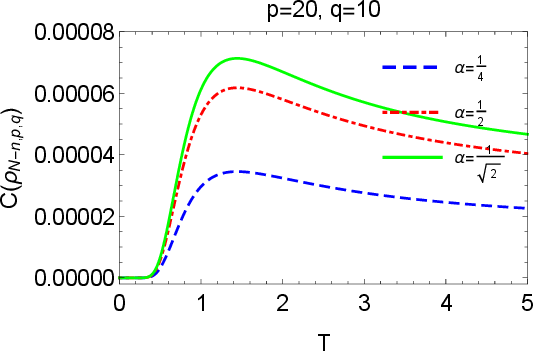}

\includegraphics[height=1.8in,width=2.0in]{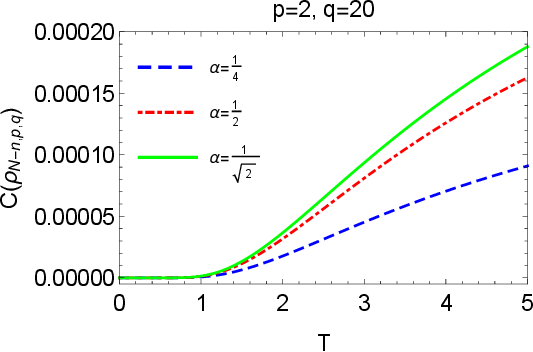}
\includegraphics[height=1.8in,width=2.0in]{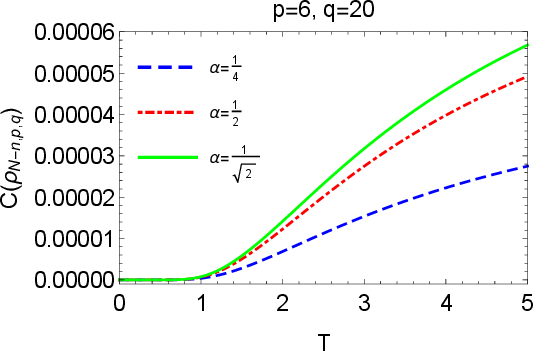}
\includegraphics[height=1.8in,width=2.0in]{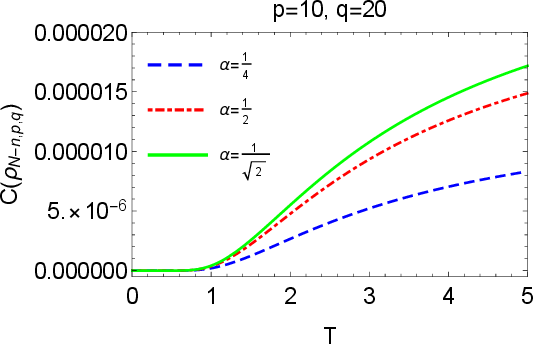}

\includegraphics[height=1.8in,width=2.0in]{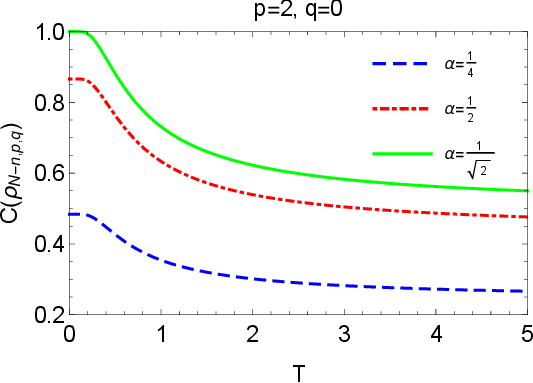}
\includegraphics[height=1.8in,width=2.0in]{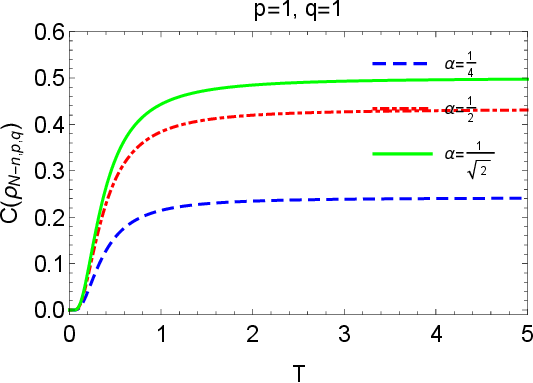}
\includegraphics[height=1.8in,width=2.0in]{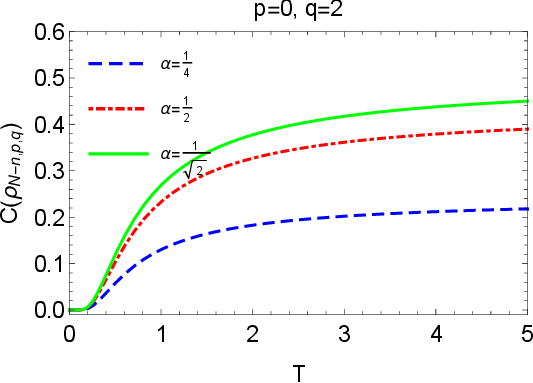}
\caption{ The genuine N-partite entanglement $C(\rho_{N-n,p,q})$ as functions of  the Hawking temperature $T$ for different $p$, $q$, and $\alpha$,  where $\omega=1$. }\label{Fig1}
\end{figure}

In Fig.\ref{Fig1}, we plot the genuine N-partite entanglement $C(\rho_{N-n,p,q})$ as functions of the Hawking temperature $T$ for different $p$, $q$, and $\alpha$.
From  the first line of Fig.\ref{Fig1}, we can see that the physically accessible genuine N-partite entanglement of Dirac field decreases to a fixed value with the Hawking temperature $T$, while quantum entanglement of bosonic field vanishes at the limit of an extreme  black hole \cite{L40}.
For the extreme black hole ($T\rightarrow \infty$), we find from Eq.(\ref{Q27}) that
$$\lim_{T\rightarrow \infty}C(\rho_{N-n,p,q})=2\alpha\sqrt{1-\alpha^2}(\frac{1}{\sqrt{2}})^n.$$
From the second and third lines of Fig.\ref{Fig1}, we can also see that
the relative numbers of the accessible to the inaccessible modes leads to a monotonic or non-monotonic increase in the physically inaccessible genuine N-partite entanglement with the Hawking temperature. In other words, the inaccessible  entanglement relies on
a decreasing function $\frac{1}{\sqrt{e^{-\frac{\omega}{T}}+1}}$ and an increasing function $\frac{1}{\sqrt{e^{\frac{\omega}{T}}+1}}$, resulting in monotonic or non-monotonic increases with the Hawking temperature. For $p>q$ (means $p>n/2$ and $q<n/2$), i.e., the number of modes outside event horizon $p$ is larger than that inside event horizon $q$,
$C(\rho_{N-n,p,q})$ is non-monotonic, such as $\{p=20,q=2\}$, $\{p=20,q=6\}$, and $\{p=20,q=10\}$ in the second line of Fig.\ref{Fig1}.  The maximum of general inaccessible entanglement  $C(\rho_{N-n,p,q})$ is $\rm{exp}(-\frac{\omega}{T})=\frac{q}{p}$. Oppositely, for $p\leq q$, the inaccessible genuine N-partite entanglement $C(\rho_{N-n,p,q})$ is monotonic,  such as $\{p=2,q=20\}$, $\{p=6,q=20\}$, and $\{p=10,q=20\}$  in the third line of Fig.\ref{Fig1}. It should be pointed out that if $N\leq 3$, then the non-monotonic phenomenon of genuine entanglement never occurs. For example, if $N=3$ and  $N-n=1$, we have three  cases: (1)  $p=2$ and $q=0$; (2)  $p=q=1$; (3)  $p=0$ and $q=2$. Please refer to  the fourth line of Fig.\ref{Fig1}. For the case of $N=2$, we obtain $C(\rho_{1,1,0})=\frac{2\alpha\sqrt{1-\alpha^2}}{\sqrt{{e^{-\frac{\omega}{T}}+1}}}$ and $C(\rho_{1,0,1})=\frac{2\alpha\sqrt{1-\alpha^2}}{\sqrt{{e^{\frac{\omega}{T}}+1}}}$.
It is obvious that the bipartite entanglement is monotonic.
Therefore, the result is consistent with the previous works \cite{L48,L49,L50,L51,L52,L53,L54,L55,L56,L57}. In addition, the general entanglement depends on the initial parameter $\alpha$.

Through the above analysis, we know that the Hawking effect of the black hole redistributes the genuine N-partite entanglement, showing that the Hawking effect reduces the accessible genuine N-partite entanglement and enhances monotonically or non-monotonically the inaccessible genuine N-partite entanglement. Naturally, we want to ask: do they satisfy the restrictive relationship? The answer is yes. Through detailed calculations from Eq.(\ref{Q27}), we can obtain two restrictive relationships between physically accessible and inaccessible genuine N-partite entanglement,
which are called the restrictive monogamy of genuine N-partite entanglement. The first restrictive monogamy of genuine N-partite entanglement  in Schwarzschild spacetime reads
\begin{eqnarray}\label{Q29}
\sum_{p=0}^nC_n^p C^2(\rho_{N-n,p,q})=\sum_{p=0}^n4C_n^p\frac{\alpha^2({1-\alpha^2})}{({e^{-\frac{\omega}{T}}+1})^{p}({e^{\frac{\omega}{T}}+1})^q} =\big(2\alpha\sqrt{{1-\alpha^2}}\big)^2,
\end{eqnarray}
where the binomial coefficient is defined by $C_n^p=\frac{n!}{p!(n-p)!}$ and $2\alpha\sqrt{{1-\alpha^2}}$ is the initial entanglement of Eq.(\ref{Q14}).
The second restrictive monogamy of entanglement  becomes
\begin{eqnarray}\label{Q30}
\sum_{q=0}^{\frac{n}{2}}C_{\frac{n}{2}}^q C(\rho_{N-n,p,q})=2\alpha\sqrt{{1-\alpha^2}},
\end{eqnarray}
where $n$ must be an even number. Eqs.(\ref{Q29}) and (\ref{Q30}) provide the restrictive relationships between accessible and inaccessible genuine entanglement in the N-partite system, indicating that when one of the entanglements decreases, the remaining entanglements must increase. In other words, the restrictive monogamy controls how quantum information of the black hole travels the interior region and the exterior region. Note that when $n$ increases, since  $\sum_{p=0}^nC_n^p$ or $\sum_{q=0}^{\frac{n}{2}}C_{\frac{n}{2}}^q$ increases, every entanglement $C(\rho_{N-n,p,q})$ must decrease due to the restrictive relations of Eqs.(\ref{Q29}) and (\ref{Q30}). This is an alternative explanation for the observed decrease in genuine entanglement with the increase of the  $n$ in Fig.\ref{Fig1}.

\section{ Conclusions  \label{GSCDGE}}
The Hawking effect on the genuine N-partite entanglement of fermionic  fields
in the Schwarzschild black hole has been investigated. We initially assume that $N$ observers share an N-partite entangled state of fermionic  fields in the asymptotically flat region.
Next, we consider that $n$ ($n<N$) observers hover near the event horizon of the Schwarzschild black hole, and $N-n$ observers remain stationary in the asymptotically flat region. Interestingly, we obtain a concise analytic expression including all physically accessible and inaccessible genuine entanglement in the N-partite system. It is shown that physically accessible genuine N-partite entanglement of fermionic  fields decreases to a fixed value with the Hawking temperature, while quantum entanglement of bosonic fields vanishes in the case of an extreme black hole. In addition, the physically inaccessible genuine N-partite entanglement increases either monotonically or non-monotonically  with the Hawking temperature, which relies on  the ratio of the inaccessible to the accessible number of modes, showing that
the result is different from the behavior of  the inaccessible bipartite and tripartite entanglement  in curved spacetime \cite{L48,L49,L50,L51,L52,L53,L54,L55,L56,L57}. The maximum for the non-monotonic genuine N-partite entanglement equals  $\rm{exp}(-\frac{\omega}{T})=\frac{q}{p}$ with accessible $p$ modes and inaccessible $q$ modes. Finally, we obtain two restrictive relationships between the accessible and inaccessible genuine N-partite entanglement, meaning that quantum information of the black hole satisfies the law. This conclusion provides a more comprehensive insight into the Hawking effect of the black hole.

\appendix
\onecolumngrid
\section{Sub-matrixes $\mathcal{M}$, $\mathcal{N}$, and $\mathcal{V}$ }
The sub-matrixes $\mathcal{M}$, $\mathcal{N}$, and $\mathcal{V}$ are $2^n\times2^n$ dimensions. By analyzing  Eqs.(\ref{Q22}) and (\ref{Q23}), we find that the basis set for sub-matrixes $\mathcal{M}$ corresponding to $\boldsymbol{\rho_M}$ is $\{|\bar00...00\rangle, \ldots, |\bar01...11\rangle\}$, where the base corresponding to the element $\frac{\alpha^2}{({e^{-\frac{\omega}{T}}+1})^{n-i}({e^{\frac{\omega}{T}}+1})^i}$ include $i$ ``1",
\begin{eqnarray}
\nonumber &&|\bar00...00\rangle\langle\bar00...00|:\frac{\alpha^2}{({e^{-\frac{\omega}{T}}+1})^n},\quad |\bar00...01\rangle\langle\bar00...01|:\frac{\alpha^2}{({e^{-\frac{\omega}{T}}+1})^{n-1}({e^{\frac{\omega}{T}}+1})}, \\\nonumber
&&|\bar00...010\rangle\langle\bar00...010|:\frac{\alpha^2}{({e^{-\frac{\omega}{T}}+1})^{n-1}({e^{\frac{\omega}{T}}+1})},\quad |\bar00...011\rangle\langle\bar00...011|:\frac{\alpha^2}{({e^{-\frac{\omega}{T}}+1})^{n-2}({e^{\frac{\omega}{T}}+1})^2},\\\nonumber &&|\bar00...0100\rangle\langle\bar00...0100|:\frac{\alpha^2}{({e^{-\frac{\omega}{T}}+1})^{n-1}({e^{\frac{\omega}{T}}+1})}, \quad ...,\\\nonumber
&&|\bar01...10\rangle\langle\bar01...10|:\frac{\alpha^2}{({e^{-\frac{\omega}{T}}+1})({e^{\frac{\omega}{T}}+1})^{n-1}},\quad |\bar01...11\rangle\langle\bar01...11|:\frac{\alpha^2}{({e^{\frac{\omega}{T}}+1})^n}.
\end{eqnarray}
The sub-matrix $\mathcal{M}$ can be specifically expressed as
\begin{eqnarray}\label{QQQQ24}
\mathcal{M}=\alpha^2 \left(\!\!\begin{array}{cccc}
\frac{1}{({e^{-\frac{\omega}{T}}+1})^n} &  & &\\
  & \frac{1}{({e^{-\frac{\omega}{T}}+1})^{n-1}({e^{\frac{\omega}{T}}+1})}& &\\
   &  & \ddots&\\
  &  & &\frac{1}{({e^{\frac{\omega}{T}}+1})^n} \\
 \end{array}\!\!\right).
\end{eqnarray}
Here, the sub-matrix $\mathcal{N}$ corresponding to $\boldsymbol{\rho_N}$ is given by
\begin{eqnarray}\label{Q25}
\mathcal{N}= \left(\!\!\begin{array}{ccccccc}
 0 &  & & & & & \\
  &  \ddots& & & & & \\
  &  & 0& & & & \\
  &  & & 1-\alpha^2& & & \\
  &  & & &0 & & \\
  &  & & & &\ddots & \\
  &  & & & & & 0\\
 \end{array}\!\!\right),
\end{eqnarray}
where  the element $1-\alpha^2$ is fixed in the position ($2^{\rm p},2^{\rm p}$), and the sub-matrix  $\mathcal{V}$ corresponding to $\boldsymbol{\rho_V}$ reads
\begin{eqnarray}\label{Q26}
\mathcal{V}= \left(\!\!\begin{array}{ccccccc}
  &  & & & & &0 \\
  &  & & & & \udots& \\
  &  & & & 0& & \\
  &  & & \frac{\alpha\sqrt{1-\alpha^2}}{\sqrt{({e^{-\frac{\omega}{T}}+1})^{p}({e^{\frac{\omega}{T}}+1})^q}} & & & \\
  &  & 0& & & & \\
  &  \udots& & & & & \\
  0&  & & & & & \\
 \end{array}\!\!\right),
\end{eqnarray}
where the element $\frac{\alpha\sqrt{1-\alpha^2}}{({e^{-\frac{\omega}{T}}+1})^{p}({e^{\frac{\omega}{T}}+1})^q}$ is fixed in the position ($2^{\rm q},2^{\rm p}$).

\begin{acknowledgments}
S.M. Wu was supported by the National Natural
Science Foundation of China (12205133) and LJKQZ20222315. T. Liu. was supported by National Natural Science
Foundation of China (12203009); Chutian
Scholars Program in Hubei Province (X2023007); Hubei
Province Foreign Expert Project (2023DJC040).
J. Wang was supported by the National Natural Science
Foundation of China (12122504).
\end{acknowledgments}



\begin{thebibliography}{99}
\bibitem{L24}
B. P. Abbott $et$ $al.$, Phys. Rev. Lett. {\bf116}, 061102 (2016).

\bibitem{L25}
The Event Horizon Telescope Collaboration, Astrophys. J. Lett. {\bf875}, L1 (2019).

\bibitem{L26}
 The Event Horizon Telescope Collaboration, Astrophys. J. Lett. {\bf875}, L2 (2019).

\bibitem{L27}
The Event Horizon Telescope Collaboration, Astrophys. J. Lett. {\bf875}, L3 (2019).

\bibitem{L28}
The Event Horizon Telescope Collaboration, Astrophys. J. Lett. {\bf875}, L4 (2019).

\bibitem{L29}
The Event Horizon Telescope Collaboration, Astrophys. J. Lett. {\bf875}, L5 (2019).

\bibitem{L30}
The Event Horizon Telescope Collaboration, Astrophys. J. Lett. {\bf875}, L6 (2019).

\bibitem{L31}
Event Horizon Telescope Collaboration, Astrophys. J. Lett. {\bf930}, L17 (2022).

\bibitem{L32}
T. Damour, R. Ruffini, Phys. Rev. D {\bf14}, 332 (1976).

\bibitem{L33}
L. Bombelli, R. K. Koul, J. Lee, and R. D. Sorkin, Phys.
Rev. D 34, 373 (1986).

\bibitem{L34}
S. W. Hawking, Commun. Math. Phys. 43, 199 (1975).

\bibitem{L35}
S. W. Hawking, Phys. Rev. D 14, 2460 (1976).

\bibitem{L38}
I. Fuentes-Schuller, and R. B. Mann, Phys. Rev. Lett. {\bf 95},120404 (2005).

\bibitem{L39}
G. Adesso, I. Fuentes-Schuller, and M. Ericsson, Phys. Rev. A {\bf76}, 062112 (2007).

\bibitem{L40}
Q. Pan, J. Jing, Phys. Rev. D {\bf78}, 065015 (2008).

\bibitem{L41}
D. E. Bruschi, J. Louko, E. Mart\'{\i}n-Mart\'{\i}nez, A. Dragan, and I. Fuentes,
Phys. Rev. A {\bf82}, 042332 (2010).

\bibitem{L42}
E. Mart\'{\i}n-Mart\'{\i}nez, L. J. Garay, and J. Le\'{o}n, Phys. Rev. D
{\bf82}, 064006 (2010).

\bibitem{L43}
B. N. Esfahani, M. Shamirzaie, and M. Soltani, Phys. Rev. D {\bf84}, 025024 (2011).

\bibitem{L44}
D. E. Bruschi, A. Dragan, I. Fuentes, and J. Louko, Phys. Rev. D {\bf86}, 025026 (2012).

\bibitem{L45}
M. R. Hwang, D. Park, and E. Jung, Phys. Rev. A {\bf83}, 012111 (2011).

\bibitem{L46}
S. M. Wu, H. S. Zeng, Eur. Phys. J. C {\bf82}, 716 (2022).

\bibitem{L47}
T. Liu, J. Jing, J. Wang, Adv. Quantum Technol. {\bf1}, 1800072 (2018).

\bibitem{L48}
P. M. Alsing, I. Fuentes-Schuller, R. B. Mann, and T. E.
Tessier, Phys. Rev. A {\bf74}, 032326 (2006).

\bibitem{L49}
J. Wang, Q. Pan, J. Jing, Phys. Lett. B {\bf692}, 202  (2010).

\bibitem{L50}
J. Wang, J. Jing, Phys. Rev. A {\bf83}, 022314  (2011).

\bibitem{L51}
S. Moradi, Phys. Rev. A {\bf79}, 064301 (2009).

\bibitem{L52}
E. Mart\'{\i}n-Mart\'{\i}nez, I. Fuentes,  Phys. Rev. A {\bf83}, 052306 (2011).

\bibitem{L53}
J. Chang, Y. Kwon, Phys. Rev. A {\bf85}, 032302 (2012).

\bibitem{L54}
J. He, S. Xu, L. Ye,  Phys. Lett. B {\bf756}, 278 (2016).

\bibitem{L55}
W. C. Qiang, G. H. Sun, Q. Dong, and S. H. Dong,  Phys. Rev. A {\bf98}, 022320 (2018).

\bibitem{L56}
S. Xu, X. k. Song, J. d. Shi, and L. Ye, Phys. Rev. D {\bf89}, 065022 (2014).

\bibitem{L57}
A. J. Torres-Arenasa, Q. Dong, G. H. Sun, W. C. Qiang, S. H. Dong, Phys. Lett. B {\bf789}, 93 (2019).

\bibitem{QLQ57}
S. M. Wu, H. S. Zeng, Eur. Phys. J. C {\bf82}, 4 (2022).

\bibitem{L80}
S. Bhattacharya, N, Joshi,  Phys. Rev. D {\bf105}, 065007 (2022).

\bibitem{A1}
P. D. Nation, M. P. Blencowe, A. J. Rimberg, and E. Buks
Phys. Rev. Lett. {\bf103}, 087004 (2009).

\bibitem{A2}
J. Drori, Y. Rosenberg, D. Bermudez, Y. Silberberg, and U. Leonhardt,
Phys. Rev. Lett. {\bf122}, 010404 (2019).

\bibitem{A3}
M. Isoard and N. Pavloff
Phys. Rev. Lett. {\bf124}, 060401  (2020).

\bibitem{A4}
Z. Tian, L. Wu, L. Zhang, J. Jing, J. Du, Phys. Rev. D {\bf106}, L061701 (2022).

\bibitem{A5}
Y. H. Shi, $et$ $al.$, Nature Communications {\bf14}, 3263 (2023).

\bibitem{A6}
P. Xu, $et$ $al.$,  Science {\bf366}, 132 (2019).


\bibitem{Q56}
J. Jing, Phys. Rev. D {\bf70}, 065004 (2004).

\bibitem{Q57}
J. Wang, Q. Pan, J. Jing, Annals Phys. {\bf325}, 1190 (2010).


\bibitem{L14}
H. J. Briegel, D. E. Browne, W. D\"{u}r, R. Raussendorf, and M.
Van den Nest, Nat. Phys. {\bf5}, 19 (2009).




\bibitem{L16}
R. Cleve, D. Gottesman, and H. K. Lo, Phys. Rev. Lett. {\bf83}, 648
(1999).


\bibitem{L17}
A. S. S{\o}rensen and K. M{\o}lmer, Phys. Rev. Lett. {\bf86}, 4431 (2001).


\bibitem{L18}
R. Raussendorf and H. J. Briegel, Phys. Rev. Lett. {\bf86}, 5188
(2001).

\bibitem{L19}
M. Murao, D. Jonathan, M. B. Plenio, and V. Vedral, Phys. Rev.
A {\bf59}, 156 (1999).


\bibitem{L20}
M. Hillery, V. Bu\v{z}ek, and A. Berthiaume, Phys. Rev. A {\bf59}, 1829
(1999).


\bibitem{L21}
V. Scarani and N. Gisin, Phys. Rev. Lett. {\bf87}, 117901 (2001).

\bibitem{L22}
Z. Zhao, Y. A. Chen, A. N. Zhang, T. Yang, H. J. Briegel, and
J. W. Pan, Nature (London) {\bf430}, 54 (2004).

\bibitem{L23}
Y. Yeo and W. K. Chua, Phys. Rev. Lett. {\bf96}, 060502 (2006).


\bibitem{Q58}
S. M. Hashemi Rafsanjani, M. Huber, C. J. Broadbent, J. H. Eberly,
Phys. Rev. A {\bf86}, 062303 (2012).




\end{thebibliography}
\end{document}